\documentclass{elsart3} \usepackage{epsfig}

\begin{document}


\begin{frontmatter}
\title{The Large Analog Bandwidth Recorder and Digitizer with Ordered Readout
(LABRADOR) ASIC}
\author[UH]{G.S.~Varner\corauthref{varner}}
\corauth[varner]{Corresponding author. Tel./fax: +001 808-956-2987.}
\ead{varner@phys.hawaii.edu}
\author[UH]{L.L.~Ruckman},
\author[UCI]{J.W.~Nam},
\author[UCL]{R.J.~Nichol}
\author[BA]{J.~Cao},
\author[UH]{P.W.~Gorham} and 
\author[BB]{M.~Wilcox}


\address[UH]{Department of Physics and Astronomy, University of
  Hawaii, 2505 Correa Road, Honolulu HI 96822, USA}

\address[UCI]{Department of Physics and Astronomy, University of California at
  Irvine, Frederick Reines Hall, Irvine CA 92697, USA}

\address[UCL]{Department of Physics and Astronomy, University College London,
  London WC1E 6BT, UK}

\address[BA]{was  at the Univ. of Hawaii, now with NeuroPace Inc., 1375 Shorebird Way,
  Mountain View CA 94043, USA} 

\address[BB]{was at the Univ. of Hawaii, now with Oceanit Laboratories
Inc., 9565 Kaumualii Hwy., Waimea HI 96769, USA}


\begin{abstract}
Three generations of full-custom analog integrated circuits designed
for low-power, high-speed sampling of Radio-Frequency (RF) transients
in excess of the Nyquist minimum have been developed.  These 0.25$\mu
m$ CMOS devices are denoted the Large Analog Bandwidth Recorder and
Digitizer with Ordered Readout (LABRADOR) ASICs and finally consist of
9 channels of 260 deep sampling.  Continuous sampling is provided with
common stop capability.  Input analog bandwidth is approximately 1GHz
and sampling speeds are adjustable from 0.02 to 3.7GSa/s.  Completely
parallel internal conversion supports 12-bit digitization and readout
of all 2340 cells in under 50$\mu s$.

\end{abstract}


\end{frontmatter}


\section{Introduction}
Observation of the early universe through neutrino messengers of the
highest possible energies requires a detector of enormous instrumented
volume.  One promising means to observe such a large,
radio-transparent target is viewing the Antarctic ice shelf via high
altitude balloon~\cite{ANITA}.  Such a balloon-borne detector needs
hundreds of high-speed sampling channels (multi-event buffering),
operating over a frequency band from 200-1200 MHz~\cite{SNIC}.  Since all power
must come from solar panels, and heat dissipation is a major problem,
commercial flash ADCs were precluded.

For at least two decades a number of Switched Capacitor Array (SCA)
devices have been reported in the high energy physics literature, for
example~\cite{lbl,lee,haller}, and many with sampling speeds high enough
for greater than Nyquist sampling of a GHz analog bandwidth signal.
These GSa/s devices have been used for low and high energy neutrino
detection~\cite{Kleinfelder}, particle physics~\cite{CB,Stefan} and
gamma-ray astronomy~\cite{Hess}.  However, despite such high sampling
speeds, all of these devices have analog bandwidth cutoffs which limit
their use at UHF frequencies and above.

We present here the results of three generations of a high analog
bandwidth ASIC designed to meet these instrumentation needs.


\section{Architecture}

A number of different CMOS SCA architectures have been discussed in
the literature.  An excellent summary of the storage circuit details
and performance may be found in Ref. \cite{arch_sum}.  As will be seen
below, in order to couple in high analog bandwidth it is necessary to
limit the parasitic and storage capacitance of the SCA array.  Thus a
compact, minimal storage array was considered and initial prototyping
looked promising~\cite{STRAW}.  This choice of a compact storage
matrix was guided and synergistic with very similar storage
architectures being explored for Monolithic Active Pixel Sensors
(MAPS) \cite{CAPs} for charged particle tracking.

\subsection{Theory of Operation}

Employment of SCA techniques in CMOS processes have been effective in
the areas of basic signal processing, continuous filter design, and
programmable capacitor arrays, used for Digital-to-Analog (DAC) and
Analog-to-Digital (ADC) conversion.
As elements of a basic programmable filter, a simple inline capacitor
between two switches may be used to form a frequency-controlled
resistor, with resistance $R$ given by ~\cite{allen}:

\begin{equation}
R = {1\over f_cC}
\end{equation}

for a given capacitor $C$, being switched at frequency $f_c$ [Hz].
Almost arbitrarily complex filters, composed of these variable $R$ and
$C$ configurations, can be formed and expressed in terms of poles and
zeros in a transfer function, the mathematics of which is conveniently
described via the $z$-Transform~\cite{unbehauen}, a staple of modern
signal processing.  As an example, from these simple building blocks,
first order filters can be constructed as represented by the transfer
function:
\begin{equation}
H(z) = K{{b_0 + b_1 z^{-1}}\over 1 + a_1 z^{-1}}
\end{equation}
where $z^{-1} = e^{-i\omega T}$, $T=1/f_c$, $-1 < a_1 < 1$, and $K$ is
an overall normalization constant.  Through choice of constants, one can
form a Low-pass filter ($b_0 - b_1 = 0$):
\begin{equation}
H(z) = K{{1 + z^{-1}}\over 1 + a_1 z^{-1}}
\end{equation}
or a High-pass filter ($b_0 = -b_1$):
\begin{equation}
H(z) = K{{1 - z^{-1}}\over 1 + a_1 z^{-1}}
\end{equation}
Since first-order filters only have one real pole, they cannot
directly realize band-pass or notch filters.  More flexible and
universal are filters of second-order and beyond.  Second order SC
filters are often called biquad circuits and have may be expressed as
\begin{equation}
H(z) = K{{b_0 + b_1 z^{-1} + b_2 z^{-2}}\over 1 + a_1 z^{-1} + a_2 z^{-2}}
\end{equation}
and is analogous to the continous time case where the transfer
function may be represented by
\begin{equation}
H(s) = K_0{{s^2 + d_1s + d_0}\over s^2 + c_1 s + c_0}
\end{equation}
where as long as the sampling frequency is much higher than the
signals of interest, the approximation $z^{-1} \simeq 1-i\omega T$ may
be used.  And from this point, standard pole-zero analysis can be used.

Beyond simple synthesis of rather complex filters using standard
tools, the true power of this technique lies in pairing such SCA
processing with operational amplifiers on an integrated circuit to
achieve powerful sampling and signal manipulation capabilities.  For
instance, in analogy with an R-2R ladder topology, a multiplying DAC
may be expressed using an array of switches and capacitors with the
simple transfer function~\cite{gregorian}

\begin{equation}
H(z) = z^{-{1\over 2}}\sum_{i=1}^n 2^{-i} b_i
\end{equation}
and ignoring the half-period delay indicated by the $z^{-{1\over 2}}$,
can synethize an output voltage $v_{out}$ based upon a reference
voltage $v_{ref}$ via the expression

\begin{equation}
v_{out} = v_{ref}\sum_{i=1}^n {b_i\over 2^{i}} 
\end{equation}
with the $b_i$ being the binary-coded digital signal, precisely as
expected for a DAC.  ADC topologies are now myriad and the focus of
this paper is on a specific type - the transient waveform recorder.
In some ways this makes use of the simplest SCA structure of them all,
the Sample-and-Hold (S/H) circuit.  The great power of the papers referenced
above derives from the ever increasing speed and compactness of deep
submicron CMOS processes.

While an idealized waveform recorder is simply an array of S/H
circuits, parasitic capacitances require consideration of parasitic
circuits like those referenced above.  For the specific application at
hand, parasitic inductances and capacitances are critical to storing
analog waveforms with frequency content in the Giga-Hertz range.

\subsection{Bandwidth Limitations}

In order for an SCA storage device to be useful, it must have a decent
number of storage cells.  Load capacitance increases as a function of
the number of switches connected to the incoming signal line, as well as
the resistance-shielded storage capacitances when the switches are
closed.  For a properly coupled $50\Omega$ stripline into an
ASIC, for a purely capacitive storage array, the 3dB roll-off is
given as
\begin{equation}
f_{\rm 3dB} = {1\over 2\pi Z_0 C}.
\end{equation}
Therefore, to obtain a 3dB bandwidth of 1.2GHz, the pure capacitance
must be limited to approximately $2.65pF$.  This value is already
smaller than that of the high-ESD protection diodes ($\sim 10pF$)
provided in a standard design library used and therefore the input
protection must be modified.  A more accurate assessment of the input
coupling performance requires a refined description of the input
circuit model and will be discussed in much more architecture-specific
detail below.  In summary, to realize 1GHz of analog bandwidth with
good coupling, will use the following design principles:

\vspace{0.1in}

\begin{enumerate}
\item $50\Omega $ stripline everywhere
\item minimize input protection capacitance
\item minimize switch drain, storage capacitance
\end{enumerate}

\vspace{0.1in}

Based upon these considerations, efforts have been made to maintain a
$50\Omega $ coupling across the sampling array inside the ASIC.  As a
trade-off between storage depth and parasitic drain capacitance, 256
samples per input was chosen.  Finally, the size of the storage
capacitance was studied.

\subsection{Storage Limitations}

\begin{figure}[ht]
\vspace*{0mm}
\centerline{\psfig{file=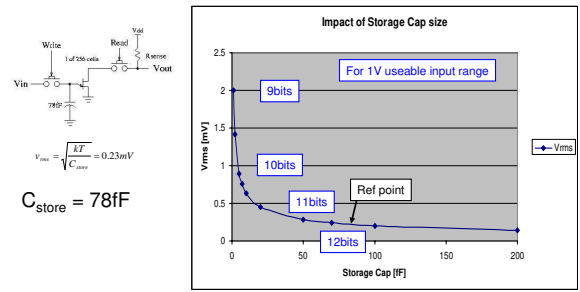,width=3.0in}}
\vspace*{0mm}
\caption{Noise limited sampling resolution as a function of storage capacitance value.}
\label{store_cap}
\end{figure}

Limits are imposed on the minimum size possible for the storage
capacitor.  Since for a S/H circuit there is no means to perform a
Correlated Double Sampling, an ambiguity in the actual stored value is
given in terms of electron counting statistics by the usual expression
\begin{equation}
v_{rms} = \sqrt{kT\over C}
\end{equation}
where $k$ is Boltzmann's constant and temperure $T$ is in Kelvin.
Matching the 12-bits of dynamic range of the sample storage conversion
to a Wilkinson ramp voltage range of about 1V, a slope of
approximately 0.25mV/least count is realized.  At this level of
sensitivity, the impact of the choice of storage capacitance value is
seen in Fig.~\ref{store_cap}.  At the upper left of this figure is a
schematic representation of the basic storage cell for the first two
generations of ASIC that utilized a transimpedance storage
architecture.  A reference value of $78fF$ is shown -- about matching
the least count of the ADC as shown.

In the last generation a different readout is employed, although the
same basic NMOS transistor gate capacitance storage is used.  This
constraint on minimum size is subsequently considered in the choice of
storage capacitance.  However reducing the storage capacitance too
much makes switch charge injection and leakage current effects more
prominent.

\clearpage

\subsection{Architectural Details}

Three generations of LABRADOR architecture ASIC have been designed,
fabricated and tested.  Their key features are summarized in
Table~\ref{lab_table}.  All have been fabricated in the TSMC $0.25\mu
$m CMOS (LO) process and have been packaged in a 100-pin plastic TQFP
package.  Economics and package performance simulations~\cite{STRAW}
drove this decision.  BGA packages were considered and may be used in
the future to reduce the contribution due to lead inductance, however
all test results are shown for this same 16.6 x 16.6 mm plastic package.

\begin{table}[htb]
\caption{\it Summary of three LABRADOR generations, where for
brevity they will be referred to by a shortened designation;
e.g. LABRADOR1 = LAB1.}
\label{lab_table}
\begin{center}
\begin{tabular}{|l|c|c|c|} \hline
{\bf Item} & {\bf LAB1} & {\bf LAB2} & {\bf LAB3} \\ \hline \hline
\# of RF inputs & 8 & 8 & 9 \\ \hline
Samples/input & 256 & 256 & 260 \\ \hline
Total samples & 2048 & 2048 & 2340 \\ \hline
\# of ADCs & 128 & 128 & 2340 \\ \hline
ADC Conversion cycles & 16 & 16 & 1 \\ \hline
Readout latency [$\mu$s]& 2200 & 2200 & $\leq 50$ \\ \hline
Analog MUX out [$\mu$s] & 25.6 & 25.6 & N/A \\ \hline
DC GND ref. & no & yes & no \\ \hline
Analog out & yes & yes & no \\ \hline
$50\Omega$ term. & end & end & input \\ \hline \hline
\end{tabular}
\end{center}
\end{table}

\begin{figure}[ht]
\vspace*{0mm}
\centerline{\psfig{file=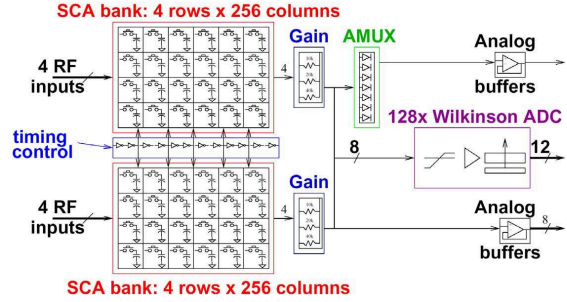,width=3.0in}}
\vspace*{0mm}
\caption{Block diagram of the LAB1/LAB2 architecture.  Samples are
stored for the 8 RF inputs in an array of 256 storage cells.  Writing
is controlled by a write pointer that continuously cycles across the
array and stored values are held upon receipt of trigger signal.
Stored values are then addressed (gain adjusted) and either stored for
conversion in an array of 128 Wilkinson ADCs or multiplexed off-chip
for external conversion.}
\label{LAB1_arch}
\end{figure}

In contrast to the first two generations of LABRADOR ASIC, the third
generation was a purely digital output device, changed input
termination scheme to be at the input, and went to a massive array of
Wilkinson ADCs (one per pixel).  These differences and lessons learned
will be highlighted below.  The architecture of first two ASICs is
illustrated schematically in Fig.~\ref{LAB1_arch}.  Examining
Table~\ref{lab_table}, the primary difference between LAB1 and LAB2
was the attempt to provide a means to internally bias the RF inputs.
This circuit did not work well due to high resistance noise coupling.
LAB1 results are similar, though better in all cases.  Eight RF input
channels are each sampled by an array of 256 SCA storage cells.
Sampling occurs continually until a trigger signal is generated. 

\begin{figure}[ht]
\vspace*{0mm}
\centerline{\psfig{file=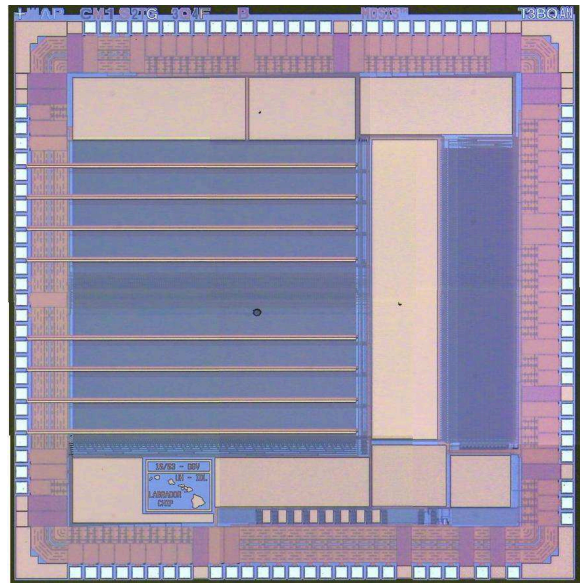,width=3.0in}}
\vspace*{0mm}
\caption{A die photograph of the LABRADOR1 ASIC.}
\label{LAB1_pic}
\end{figure}

At this point the analog samples are held and not overwritten.  These stored values are
then selected and a transimpedance relay of the stored charge is made,
which is either stored into input samples of an array of 128 channels of
Wilkinson ADC or analog multiplexed and transferred off-chip for
external ADC conversion. A die photograph of the approximately 10mm$^2$ LAB1 device is shown in
Fig.~\ref{LAB1_pic}.  

\begin{figure*}[ht]
\vspace*{0mm}
\centerline{\psfig{file=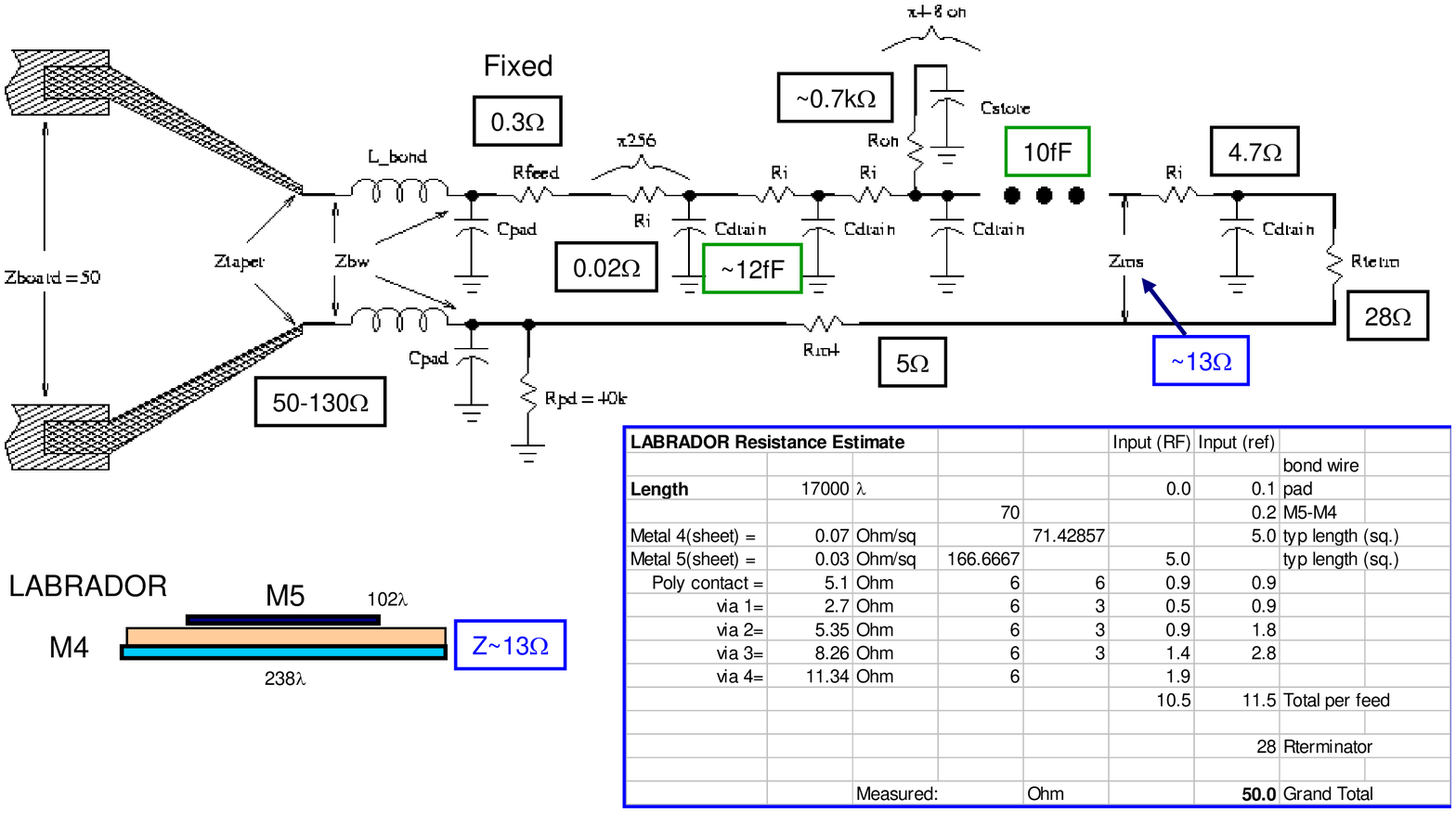,width=6.0in}}
\vspace*{0mm}
\caption{Schematic representation and resistance breakdown of the
LAB1 signal chain.  Effects due to both resistive drop across the
sampling array, as well as low impedance of the on-chip stripline, were
observed in testing.}
\label{LAB1_input}
\end{figure*}

Layout of the LAB1 ASIC quite directly follows the arrangement of the
functional blocks in the schematic diagram.  While efforts were made
to optimize the coupling of the input signal based on earlier efforts
with the STRAW~\cite{STRAW} architecture, the choice of LAB1 input
structure represented a compromise, as shown in Fig.~\ref{LAB1_input}.
Signals are straight input shots on the left and terminated in a
$28\Omega $ resistor at the right.  This choice is a trade-off between
widening the signal trace, which would lower the microstrip impedance
even below the $Z_0 = 13\Omega$ shown, or having even larger resistive
losses across the array.  These resistive losses made for a vexing
amplitude-dependence across the array.  To address this issue in LAB3,
a $50\Omega $ termination resistor is placed directly at the input to
the detector.  The termination resistor was removed from the array end.
Since offset biasing could be performed directly at this input
termination, the resistance of the signal line was unimportant and the
on-chip stripline could be made exactly $Z = 50\Omega$.  Any
reflection at the end of the array would be back-terminated, though
this stub is short.  At maximum signal frequency of 1.2GHz, for a
stripline of 2mm long (about 10ps at $v_{prop} \simeq {2\over 3}c$),
the phase introduced by this stub is about
\begin{equation}
2\cdot{10{\rm ps} \over (1.2{\rm GHz})^{-1}} \cdot (360^{\circ}) \simeq 8.6^{\circ}
\end{equation}
which is acceptable, though for operation at higher frequencies, such
effects may be non-negligible.  In all cases the input protection
diodes have been completely removed.  Current discharge is provided
through a $20k\Omega $ pull-down resistor to ground and voltage
clamping is provided by external back-to-back RF diodes.

\begin{figure*}[ht]
\vspace*{0mm}
\centerline{\psfig{file=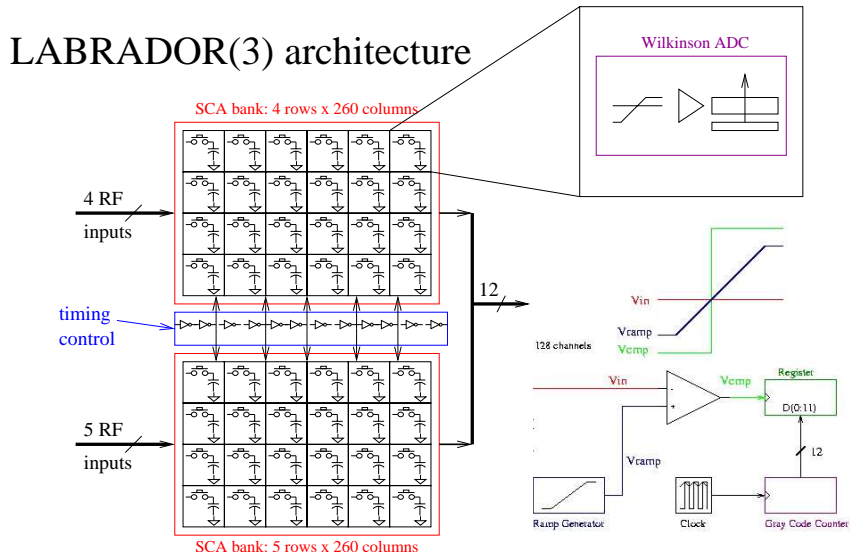,width=4.5in}}
\vspace*{0mm}
\caption{Block diagram of the LAB3 architecture.  In contrast to
the LAB1/LAB2, the stored analog signal is never transferred.
Instead, direct Wilkinson conversion is done within each storage cell.}
\label{LAB3_arch}
\end{figure*}

Other lessons gleaned from the first two LABRADOR generations included
observing that while having analog samples available for external
digitization has merits, non-linearities in the transimpedance
response and temperature dependence were major issues.  As space was
available to permit completely parallel conversion of all 9 channels
by 260 samples, in-situ conversion was adopted, as illustrated in
Fig.~\ref{LAB3_arch}. Including four extra ``tail'' samples avoids a
sampling record gap during the interval in which the write pointer is
returning to the beginning of the sampling window.

Details of the required timing and offset calibrations are discussed
below.  In order to accomodate the additional samples, as well as
provide space for a Wilkinson comparator and 12-bit latch in each
pixel storage cell, the die had to increase slightly to approximately
3.2 by 2.8 mm.  Metal fill rules required covering the interesting
parts of the die, making LAB3 far less photogenic than LAB1/LAB2
and thus not included.  Addition of a 9th channel was done to allow
insertion of a common reference clock into the data stream for each
LABRADOR.  This was found to be use for improving the temporal
alignment of waveforms recorded by different chips.

All three generations use the same write pointer structure.  This is a
classical voltage-controlled inverter chain, with an odd number of
stages such that a ripple continuously propages.  An XOR circuit and a
look-ahead signal are used to open each storage gate for the time it
takes to transition from the look ahead to current locations (4-6 samples).

Despite best efforts at balancing the threshold voltage and NMOS versus
PMOS L:W ratios, some amount of propagation variation is expected when
the ripple edge across the array is transitioning low-to-high versus
high-to-low, as shown below.

\indent The ramping voltage for Wilkinson conversion is generated by
using a current source and either an internal or external reference
capacitor.  In all testing shown below, an external 200pF capacitor is
used.  An external ($68k\Omega $) bias resistor sets the drive
strength of the current source to approximately $20\mu$A.  A common
Gray-code counter is provided on chip and broadcast to all SCA cells.
When the ramp threshold is crossed in a particular cell, the current
count value is latched.  Upon completion of ramping, all 2340 12-bit
values are available for random-access readout.\\

\subsection{Design Evolution}

In summary, the biggest changes in going from the LAB1/LAB2
architecture to the LAB3 are

\vspace{0.1in}

\begin{enumerate}
\item direct termination at array input
\item Wilkinson conversion in each storage cell (no analog signal transfer)
\item addition of a 9th (clock reference) channel
\end{enumerate}

\vspace{0.1in}

and by these choices good performance results have been obtained, as
documented below.

\clearpage


\section{Test Results}

A variety of tests have been carried out to evaluate the performance
of the LABRADOR series of waveform recorders.  These measurements
attempt to verify the degree to which the performance targets have
been met, as well as characterize the system in preparation for UHF
radio transient detection.  Because of the superior performance of
measured noise, bandwidth, linearity and digitization, results are
shown the LAB3 ASIC.

\subsection{Sampling Speed}

\begin{figure}[htb]
\vspace*{0mm}
\centerline{\psfig{file=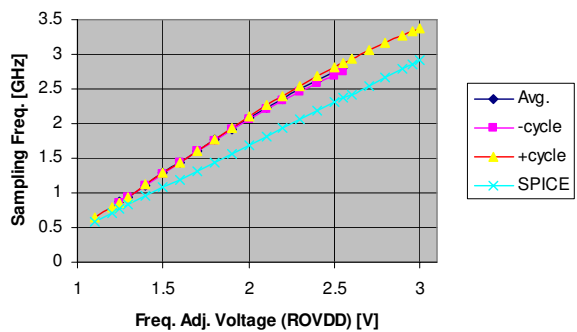,width=3.0in}}
\vspace*{0mm}
\caption{Sampling rate as a function of control voltage.  Both data
and SPICE simulation are plotted, where a difference is observed
between rising or falling edges of the ripple oscillator as described
in the text.}
\label{rovdd}
\end{figure}

The sampling speed dependence on an adjustable control voltage (ROVDD) is
plotted in Fig.~\ref{rovdd}.  Stable sampling speeds ranged from 0.02
to almost 4 GSa/s (limited by operation beyond the 2.5V nominal
VDD rail voltage).

The SPICE simulation was fairly conservative and should be considered
a lower-limit, pessimized for a worst-case spread in actual CMOS
fabrication parameter values.  While the sampling rate is defined as
the cycle average of the so-called Ripple Carry Out (RCO), which is a
copy of the write pointer monitored external to LAB3, the propagation
speed of the high-to-low and low-to-high are seen to be different.  At
a nominal 2.6GSa/s this corresponds to about a 2\% effect and is
readily calibrated out by latching the RCO bit state at the time a trigger
is recorded,
as will be discussed later.


\subsection{Input Coupling}

Pulsing the input to the LAB3 chip with a fast risetime signal, a
reflection $R = +6.8\% $ is observed.  Solving the usual expression
\begin{equation}
{Z-Z_0 \over Z+Z_0} = R
\end{equation}
an impedance value of $Z=57\Omega$ is determined.  This is consistent
with the measured $59 \Omega$ DC resistance of the fabricated device,
which appears to be about 20\% higher than specified, though within
spreads observed for silicide block in recent similar runs.

\begin{figure}[htb]
\vspace*{0mm}
\centerline{\psfig{file=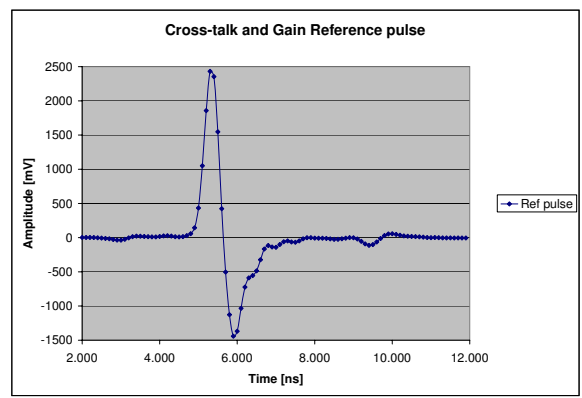,width=3.0in}}
\vspace*{0mm}
\caption{Time-domain signal of the RF pulse used to evaluate input
coupling, linearity and crosstalk, as recorded with a 3GHz bandwidth
oscilloscope.}
\label{ref_pulse}
\end{figure}

\indent Because the signal of interest is an RF signal, a standard DC
linearity scan performance is less important than evaluation with a
realistic impulsive signal.  Therefore, to determine the input
coupling, linearity and cross-talk performance, an RF impulse was used
as shown in Fig.~\ref{ref_pulse}.  Most of the signal power of
interest is in the steep high-to-low transition.

A 3GHz analog bandwidth oscillocope was used to record this reference signal.
However the signal from the pulse generator itself was not flat in
the frequency domain.  Moreover this reference pulse has been bandwidth
limited between 200-1200MHz, to match the frequency range of the ANITA
instrument signal chain, in which these measurements have been
performed.

\begin{figure}[htb]
\vspace*{0mm}
\centerline{\psfig{file=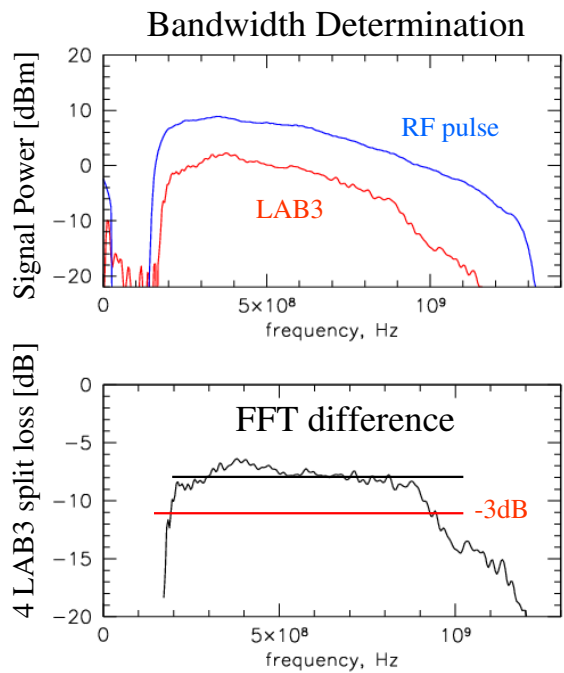,width=3.0in}}
\vspace*{0mm}
\caption{Determination of the LAB3 ASIC analog bandwidth in a test
board configuration with a four-way split of the RF signal.  In top
(blue) RF reference pulse and (red) LAB3 FFT.  At bottom is the
difference, where for perfect coupling a -6dB loss would be expected.}
\label{refp_FFT}
\end{figure}

Because determination of the analog bandwidth of the LAB3 device
requires removing the intrinsic frequency of the RF pulse itself, its
FFT has been measured and is displayed as the blue curve in Fig.~\ref{refp_FFT}.
In red in this upper plot is the recorded LAB3 response.

Taking the difference of these two curves, the analog response versus
frequency is determined and shown in the bottom plot of
Fig.~\ref{refp_FFT}.  At the left edges of the curves the impact of
the 200MHz high-pass band definition filters are seen.  Of note is
peaking of the signal in the 300-400MHz range, an effect seen in
earlier testing.  Taking the -3 dB point as the line shown, the
roll-off frequency is just over 900 MHz, though signal power is
still available out to 1200MHz.  Four LAB3 are being tested in parallel and
thus an ideal loss would be -6dB, indicating some amount of loss in
the RF signal chain and coupling into the chip.  Earlier tests on a
dedicated, single LAB3 board, without band definition filters
(e.g. 1200MHz low pass) indicated somewhat better higher frequency
response and some of this loss may be due to components on the ANITA
flight digitizer (SURF\cite{SNIC}) board used for evaluation.
Therefore this curve may be considered a conservative lower bound on
the analog bandwidth.

We note that the peaking observed is also present in the case of
gaussian noise, though the peak of the distribution is a function of
the input biasing network.  This is likely due to resonant L-C
response in the input front end and seems coupled to the cross-talk
observed below.

\subsection{Linearity}

\begin{figure}[htb]
\vspace*{0mm}
\centerline{\psfig{file=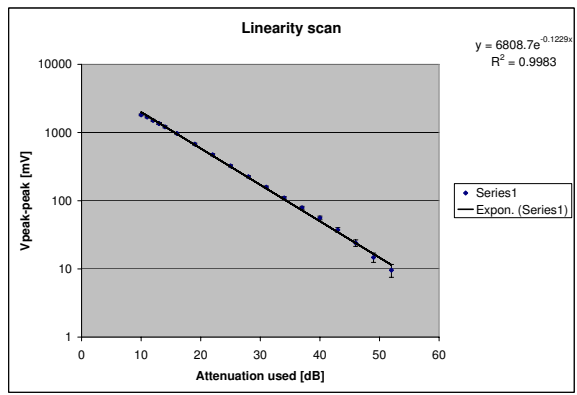,width=3.0in}}
\vspace*{0mm}
\caption{Linearity determined by attenuating an RF pulse as described in the text.}
\label{lin_scan}
\end{figure}

A determination of the linearity of the digitizing system has been
made by varying the RF signal amplitude as displayed in
Fig.~\ref{lin_scan}.

Since power attenuators are used, the response is characterized in dB
and a linear fit is observed on a logarithmic plot.  Good
linearity is seen with just a hint of saturation at large signal
amplitudes and some non-linearity at small signal amplitude due to the
coaddition of board-level noise.  Any non-linearity observed is likely
due to non-linearities in the ramp generation circuit or comparator
bias setting.  Over a span of 40dB in dynamic range, the LAB3 output
tracks input to within statistical measurement errors.

\subsection{Crosstalk}

\begin{figure}[htb]
\vspace*{0mm}
\centerline{\psfig{file=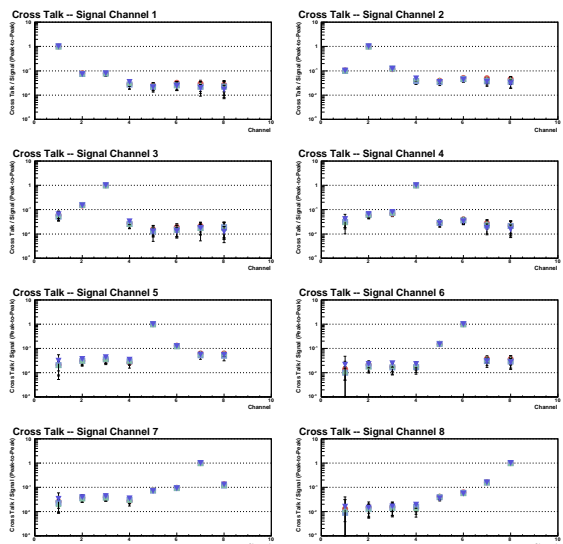,width=3.0in}}
\vspace*{0mm}
\caption{Measured crosstalk for each channel as a function of channel
into which signal is injected.  For signal in self-channel, the amplitude
is unity.  Note that these values are overestimated, as described in
the text.}
\label{crosstalk}
\end{figure}

By inserting the signal successively into the LAB3 channels, a
cross-talk correlation plot was constructed as shown in
Fig.~\ref{crosstalk}.

These values shown are determined by searching for a peak around the
time of the input signal.  Due to noise, statistically a few percent
peak is measured even for the case of no cross-talk.
Therefore the values shown are overestimated.  For RF applications,
even a 10\% voltage crosstalk is only 1\% in power. 

\begin{figure}[htb]
\vspace*{0mm}
\centerline{\psfig{file=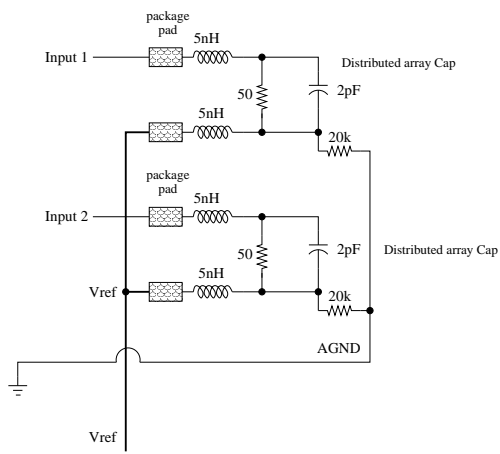,width=2.6in}}
\vspace*{0mm}
\caption{Schematic representation of the input bias circuit.}
\label{input_ckt}
\end{figure}

Nevertheless, for other applications it is important to understand the
source of this effect.  A hint to the origin of this crosstalk may be
seen in Fig.~\ref{input_ckt}.

\begin{figure}[htb]
\vspace*{0mm}
\centerline{\psfig{file=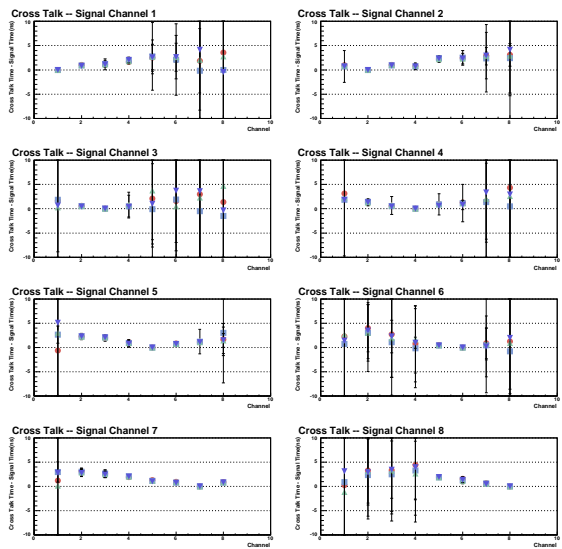,width=3.0in}}
\vspace*{0mm}
\caption{Phase lag of the measured crosstalk.  For channels separated
by the timing control section of the chip both the amplitude and phase
are less well constrained.}
\label{xtalk_phase}
\end{figure}

Similar temporal and frequency dependence to the cross-talk can be
reproduced in SPICE simulations, though the solutions are not unique.
That is, the amplitude and phase information can be mimicked by tuning
the voltage source output inductance of the pedestal network or with
respect to bond-wire inductance stray coupling.  Based on these
results, a channel-dependent phase-lag to the cross-talk was predicted
and subsequently verified qualitatively, as shown in Fig.~\ref{xtalk_phase}.

In addition, a small
component of direct radiative coupling between the on-chip striplines
cannot be ruled out, though was difficult to model (metalic heat sinks
would need to be taken properly into account in the 3D EM
simulations).  All results indicate that better packaging (lower
inductance) and stripline shielding would help improve the observed
effects.



\section{Required Calibrations and Stability}

In order to obtain the test results shown, a number of calibrations
are needed.  In the process of applying these, much improved
resolution is obtained.  Temperature dependence and timing precision
limits are considered.
  
\subsection{Gain and Pedestal Calibration}

\begin{figure}[htb]
\vspace*{0mm}
\centerline{\psfig{file=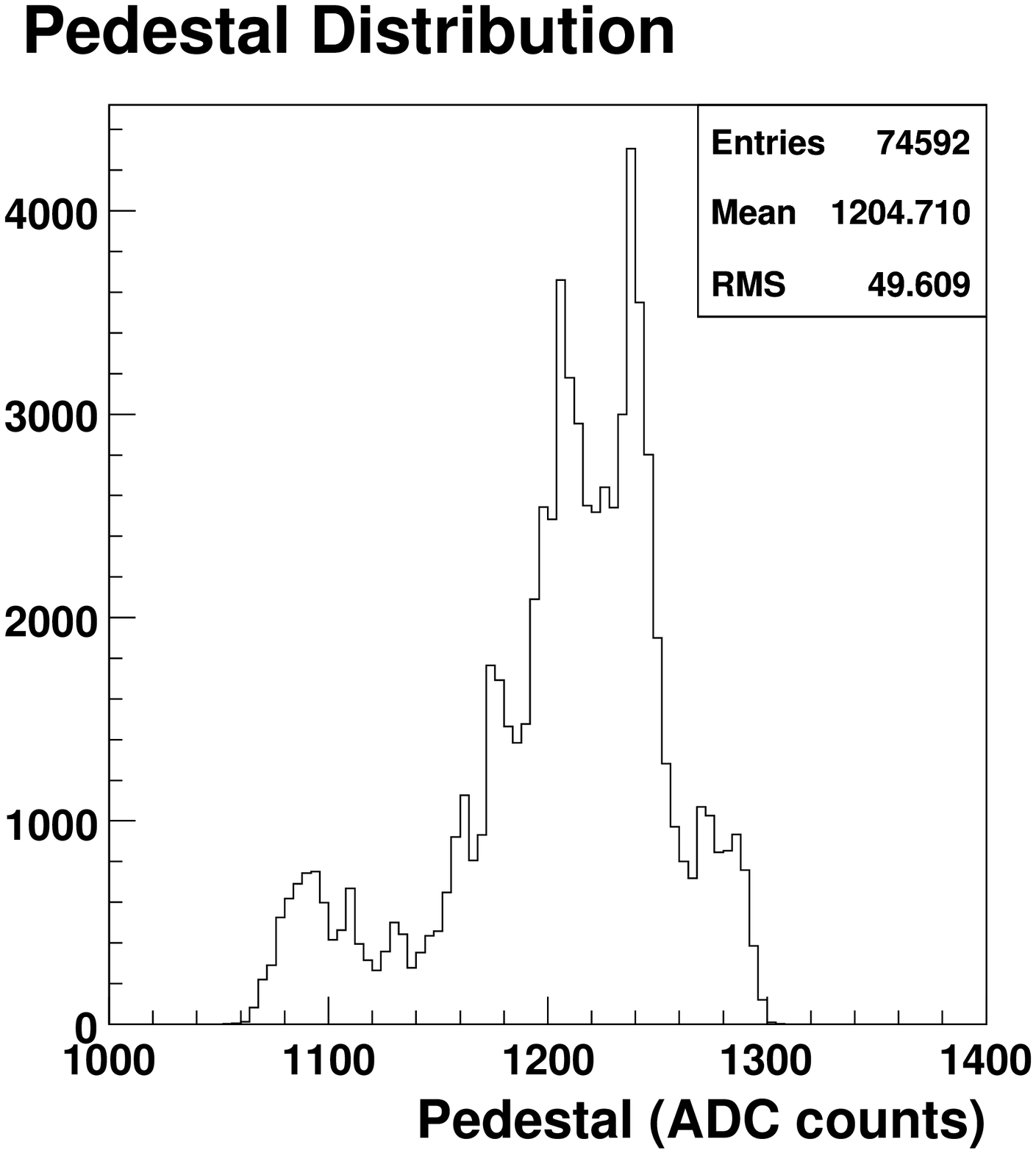,width=3.0in}}
\vspace*{0mm}
\caption{A summary of the pedestal values (in mV) for all SCAs of 36
production LAB3 tested.}
\label{all_peds}
\end{figure}

For the measurements shown, the gain has been adjusted to
approximately $1$mV/least count.  A comprehensive pedestal histogram of 
all SCA storage channels (excluding channel 9) on the 36 LAB3 flown
on ANITA is summarized in Fig.~\ref{all_peds}.

Channel 9 is excluded since it has a different voltage offset value due to
the clock input biasing.
The spread seen is a combination of 36 pedestal voltage differences,
SCA-SCA variations, and Wilkinson ramp slope and starting voltage
offsets.  Also the gain of one LAB3 (values clustered around 1100) had
an anamolously low gain.  Overall the RMS of this distibution is just over 4\%.

\subsection{Timing Calibrations}

\begin{figure}[htb]
\vspace*{0mm}
\centerline{\psfig{file=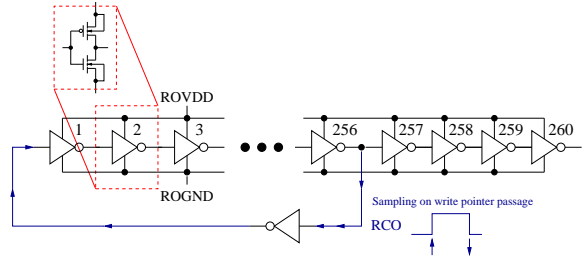,width=3.0in}}
\vspace*{0mm}
\caption{Write pointer wrap around.  While the write pointer returns
to position 0 of the array, additional tail samples are taken in order
avoid a gap in the sampling record.}
\label{ro}
\end{figure}

In order to obtain the best possible timing resolution, a number of
calibrations, due to the method in which the sampling is
implemented, must be considered.  As mentioned earlier, the write pointer is
monitored using a copy of the signal called RCO.  Since the sampling
is done in so-called Common Stop mode, it is continuous until a
trigger condition is formed.  Thus all samples have already been
recorded by the time a trigger is acted upon.  In order for sampling
to be continuous it is necessary for the write pointer to wrap around
from the end of the array to the beginning, as illustrated in Fig.~\ref{ro}.  

Four additional ``tail'' samples are provided to permit samples to be
recorded during the time in which the write pointer is returning to
the beginning of the array.  Even though the physical distance is only
20-30ps at the speed of light, the need to go through an additional
inverting stage (to form ring oscillator) and the capacitance
associated with the long signal line back to the beginning of the
array limit the speed of write pointer return.

\begin{figure*}[ht]
\vspace*{0mm}
\centerline{\psfig{file=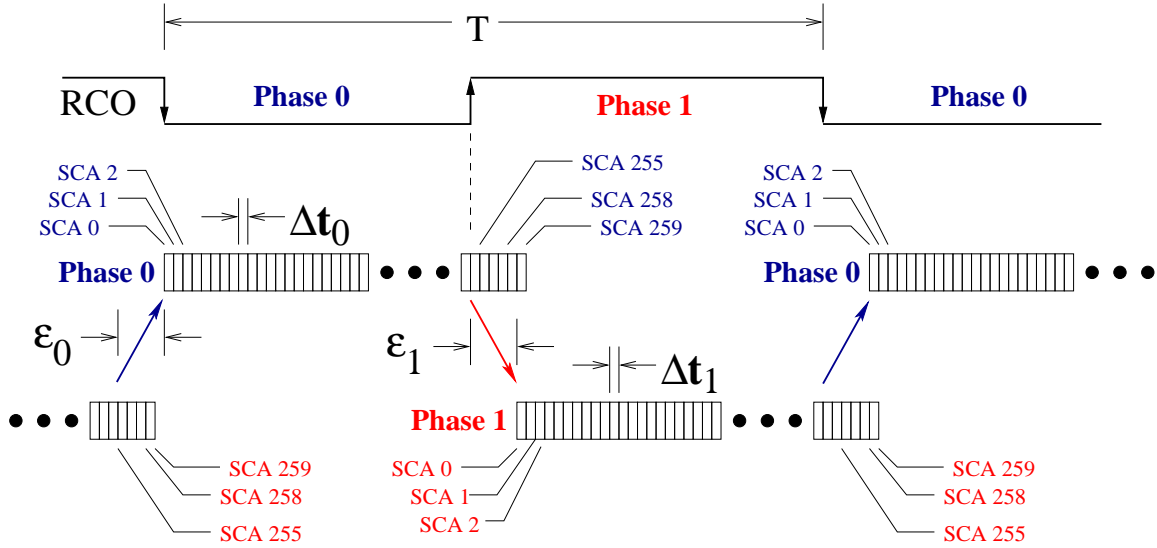,width=6.0in}}
\vspace*{0mm}
\caption{Definition of the most general LAB3 sample timing
relationships and constants.  Determination of their values is
described in the text.}
\label{calibs_all}
\end{figure*}

Also mentioned earlier, the write pointer speed of propagation across
the array is a function of the transition direction.  Likewise the
delay time of write pointer return is also RCO phase dependent.  The
most general case of these calibration constants is illustrated in
Fig.~\ref{calibs_all}.  From the measured RCO frequency ($f_{\rm RCO}$),
the sampling frequency is determined as
\begin{equation}
f_{\rm sampling} = 2 \times 256 \times f_{\rm RCO}.
\end{equation}

Expressing $f_{\rm RCO}$ in terms of its period $T$, half period
$T_0$ corresponds to RCO phase 0 and half period $T_1$ corresponds to
RCO phase 1, or
\begin{equation}
f_{\rm sampling} = 512 \times (T_0+T_1)^{-1}
\end{equation}

in which case the time step of an individual sample is expressed as 
\begin{equation}
\Delta t = {T\over 512}.
\end{equation}

In general, as mentioned, the half periods $T_0$ and $T_1$ are not
half the period $T$:
\begin{equation}
T_0 \neq T_1 \neq {T\over 2}
\end{equation}

which means that the average individual time steps in phase 0 ($\Delta
t_0$) are different from those in phase 1 ($\Delta t_1$).  Likewise
the delay time of the write point propagation for RCO $0\rightarrow 1$
($\epsilon_1$) and for RCO $1\rightarrow 0$ are in general different
and related to the difference between average $\Delta t_0$ and $\Delta
t_1$.  Finally, due to transistor threshold dispersion, the actual
widths of each of the time bins ($\Delta t_{0,1}^{0..259}$) can be
slightly different.

\begin{figure}[htb]
\vspace*{0mm}
\centerline{\psfig{file=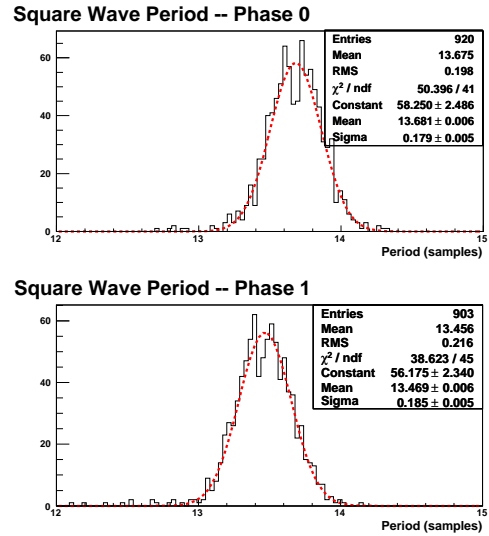,width=3.0in}}
\vspace*{0mm}
\caption{Measurement of the write pointer propagation (sampling speed)
difference for the RCO = 0 (top) and RCO = 1 (bottom) phases for a
200MHz reference clock.}
\label{RCOphase}
\end{figure}

 Using a known periodic input signal, it is possible to generate
calibration values for all of these parameters.  An example of
determination of the relative average $\Delta t_0$ and $\Delta t_1$ is
shown in Fig.~\ref{RCOphase}.

In each case the variable parameter is tuned until the spread or
offset in the determined period is minimized.  Because the period is
well determined, the procedure is very efficient and requires a relatively small
amount of calibration data.

\clearpage

\begin{figure}[htb]
\vspace*{0mm}
\centerline{\psfig{file=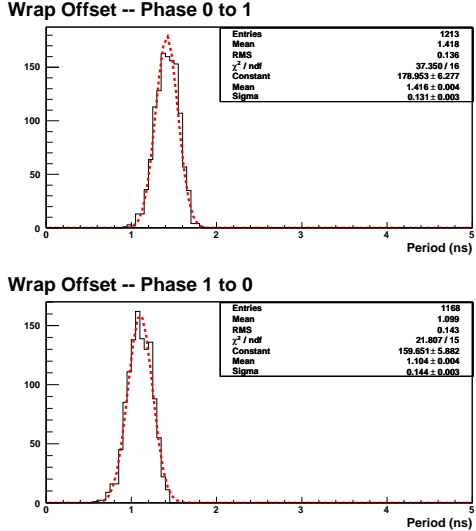,width=3.0in}}
\vspace*{0mm}
\caption{Extraction of the wrap timing offsets ($\epsilon_0$ and
$\epsilon_1$) for a given LAB3.}
\label{wrapOff}
\end{figure}

Similarly, the write pointer wrap around delays, $\epsilon_0$ and
$\epsilon_1$, may be determined by constraining the measured period to
be consistent across the write pointer wrap around.  An example is
shown in Fig.~\ref{wrapOff}.

To a certain extent these calibration steps must be bootstrapped.  For
example, correctly minimizing the error on these $\epsilon $
parameters requires that the average time steps in each of the RCO phases be
correctly determined.  A subtlety here is that the RCO phase is
recorded at the time a trigger signal (hold) is issued.  Because the
RCO latching in the data is not completely synchronous, there is in
general a delay between the measured value of RCO and its actual
value.  This ambiguity is resolved by assigning a phase delay between
the measured RCO that depends upon the address at which the hold was
issued, the so-called ``HitBus'' value.  The value of this delay is
tuned in the data until the width of the measured period is again
minimized.

\begin{figure}[htb]
\vspace*{0mm}
\centerline{\psfig{file=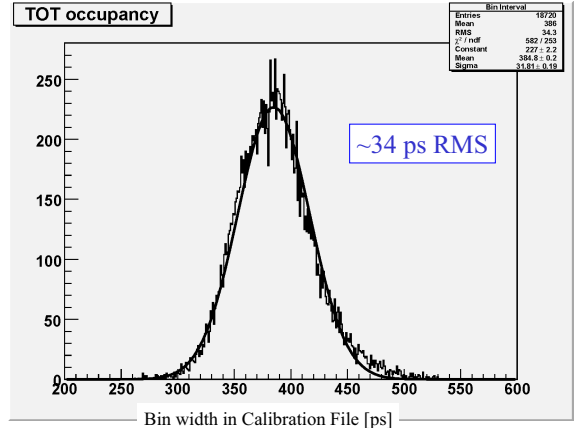,width=3.0in}}
\vspace*{0mm}
\caption{Summary distribution of the calibrated individual time bin
widths for all SCAs in 36 LAB3 ASICs.}
\label{binbybin}
\end{figure}

Finally, using a high frequency clock it is possible to constrain the
average half period and assign its average value to the $\Delta
t_{0,1}^{0..259}$ bin in which the positive/negative lobe peaks.
Using this prescription the distribution histogrammed in
Fig.~\ref{binbybin} is obtained.  

\subsection{Time Resolution Limitations}

Applying these timing corrections to the data leads to an improvement
in the time resolution of signals in the data.  The precise
improvement depends upon the signal distance within the window (cumulative
error) and method for correlating signal shapes to extract a timing
feature for comparison.  

\begin{figure}[htb]
\vspace*{0mm}
\centerline{\psfig{file=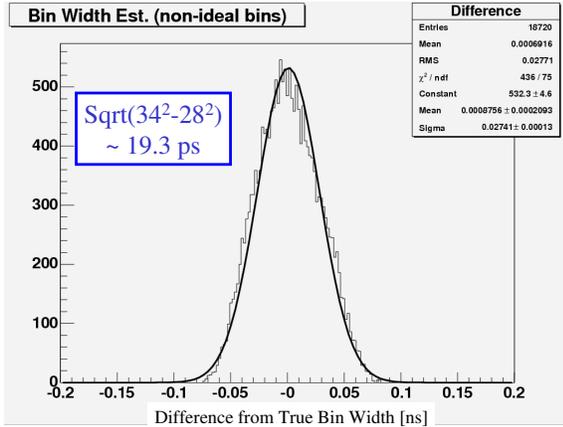,width=3.0in}}
\vspace*{0mm}
\caption{Monte carlo.}
\label{err_est}
\end{figure}

To understand the intrisic performance limits and the significance of
the bin-by-bin correction, a simple Monte Carlo study was performed to
determine the extent to which the technique used to extract the
observed timings would lead to the observed distribution.  Introducing
a completely random scatter (uniform distribution) of 15\% to the
nominal 386ps bin width, 600MHz sine MC was then synthesized and the
algorithm applied.  A value of 15\% was determined empirically to
provide a good representation of the observations in data.  Due to
irreducible errors in the specific implementation of this
zero-crossing technique, application of these constants improves the
timing resolution to about 28ps, as shown in Fig.~\ref{err_est}.  This
has improved the resolution by about 20ps in quadrature, though
perhaps there is still room for improvement.  Since two edges are used
to determine this time interval, the single edge measurement is about
$28ps/\sqrt{2}$ or about 20ps and probably is a limit with the current
LAB3.

\clearpage

These determined parameters appear to be stable in time and only
depend upon thermal effects, and described next.


\subsection{Temperature Dependence}

The voltage controlled oscillator for the write pointer is
fundamentally temperature dependent.  During operation of LAB1 an
external delay locking loop circuit was used to adjust ROVDD to
compensate.  However this circuit suffered from large phase noise as
well as a nasty habit of locking onto a frequency subharmonic at
power-on.  Therefore, with the addition of a dedicated timing channel
-- needed to precisely align multiple LAB3 waveforms offline -- ROVDD
was fixed and timebase correction is implemented by fitting to the
period of the reference clock.  

\begin{figure}[htb]
\vspace*{0mm}
\centerline{\psfig{file=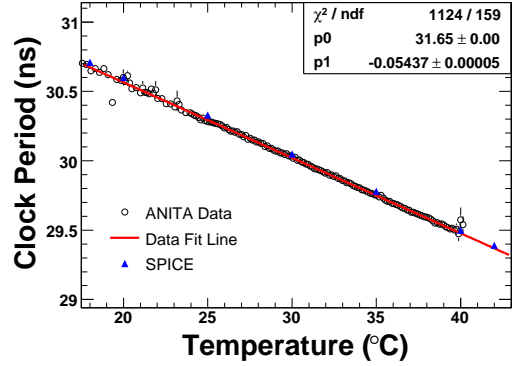,width=3.0in}}
\vspace*{0mm}
\caption{Measured and SPICE simulated temperature dependence of the
LAB3 sampling period.}
\label{RCOtempdep}
\end{figure}

The temperature dependence of the sampled frequency is shown in
Fig.~\ref{RCOtempdep}. Good agreement is seen with SPICE simulations
of the temperature dependence, once an operating reference point is
set.  This fine tuning is needed to correct for the overly pessimistic
parasitic capacitance estimate used earlier in simulating the ripple
oscillator frequency.  Using the reference clock signal on channel 9,
this temperature dependence of the VCO is corrected in the offline
analysis.  A fit to this dependence gives a change of approximately
55ps/$^{\circ }$C over the 30ns period of the 33MHz reference clock,
or about 0.2\%/$^{\circ }$C.

\begin{figure}[htb]
\vspace*{0mm}
\centerline{\psfig{file=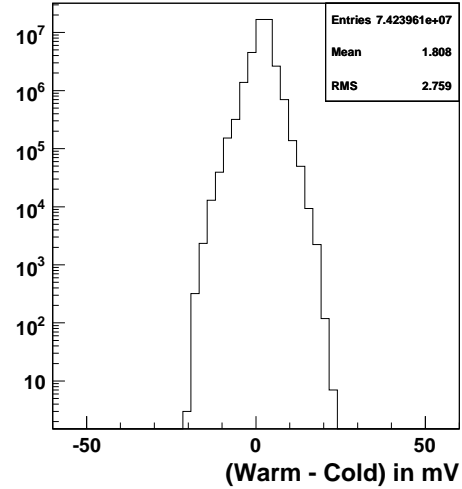,width=2.8in}}
\vspace*{0mm}
\caption{Difference in pedestal between dedicated pedestal runs taken
30 hours apart, at a difference in ambient temperature of 17 degrees
Celcius.}
\label{pedTstability}
\end{figure}

In contrast, the pedestals are a very weak function of
temperature.  In Fig.~\ref{pedTstability} is displayed the difference
in pedestal values taken after an ambient temperature change of
approximately 17$^\circ$C.

Taking this difference, 
an estimate of the pedestal temperature dependence is
\begin{equation}
{\rm PED_{avg}} = +0.052 \cdot {{\rm ADC counts}\over ^{\circ}C}
\end{equation}

\clearpage

\begin{figure}[htb]
\vspace*{0mm}
\centerline{\psfig{file=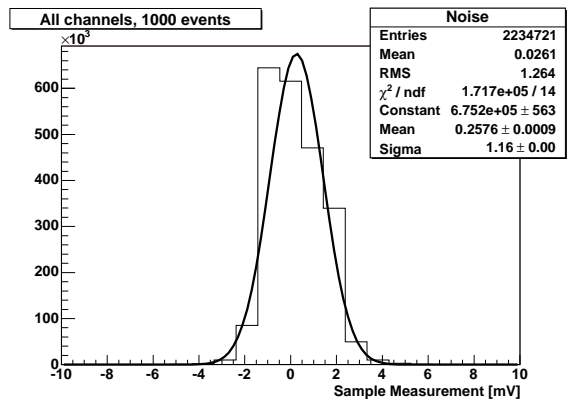,width=3.0in}}
\vspace*{0mm}
\caption{Sample LAB3 1k event noise run, with all 9 channels combined
into a single distribution.}
\label{noise}
\end{figure}

For reference, and to illustrate the typical chip-level noise, an
example noise run is shown in Fig.~\ref{noise}.  Representative noise values
are about $1.3mV_{rms}$, though there is some non-gaussian behavior in
the combined distribution of 2.2 million samples from 9 separate
RF channels.

\subsection{Interleaved Sampling}

\begin{figure}[htb]
\vspace*{0mm}
\centerline{\psfig{file=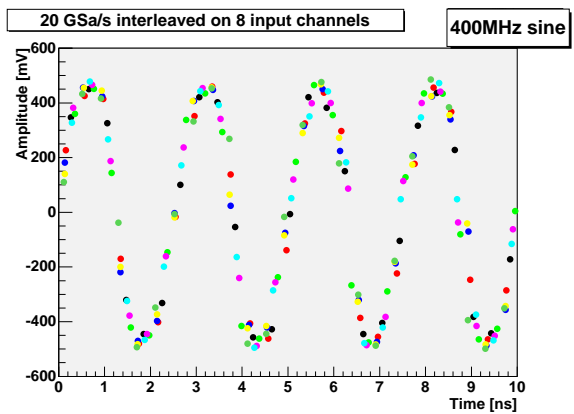,width=3.0in}}
\vspace*{0mm}
\caption{Example of 20GSa/s interleaved, single-shot waveform
recording of a 400MHz sine wave signal on 8 LAB3 input channels, each
plotted with a different color.}
\label{interleave}
\end{figure}

Right at Nyquist sampling of UHF RF sine wave signals visually appears
undersampled to most observers.  This is due to expectations from
seeing smooth curves generated by 20+ GHz offset interleaved sampling
of a repetitive waveform, provided by most digital signal
oscilloscopes.  By providing precise external delays it is possible to
enhance the sampling speed and provide oversampling with the LAB3
chip.  Interleaving of 8 inputs, running at 2.5GSa/s each has been
done to provide single-shot recording of a 400MHz sine wave signal at
20GSa/s, as shown in Fig.~\ref{interleave}.  Each color represents the
samples recorded by a single channel.

While there is some scatter due to the delays not being perfectly
tuned, this indicates that there is still more performance to be
gained by increasing the analog bandwidth yet further and implementing
such interleaving.  For low power and very high sampling rate
applications, where signals may not be repetitive, this technique may
be useful.  This and other improvements are considered next.


\section{Future Directions}
Beyond increasing analog bandwidth, to fully exploit the enhanced sampling
speed of deep submicron processes, there is a desire to increase
sampling depth.  This is being explored in a follow-on device
designated the Buffered LABRADOR (BLAB) ASIC.  While the sampling
speed increases below 0.25$\mu $m, loss of dynamic range due to
reduced rail voltages and increased leakage current may preclude
going to smaller feature sizes.


\section{Applications}
During December 2006 to January 2007, 36 LAB3 ASICs flew successfully
at 120,000 feet for 35 days around the Antarctic continent.  Some of
the test results shown above are from this data set.  During this same
period, test deployments for an in-ice radio detector, using this
device, were made at the south pole in conjunction with the IceCube
array~\cite{AURA}.  Recently, these ASICs were evaluated in a
collidering beam environment for upgrade of the Belle Time-Of-Flight
readout~\cite{TOF}, and a variant for operation at a Super
B-factory~\cite{superb} is being developed for high timing precision,
single photon recording~\cite{PROMPT}.  For these future applications,
a deeper sampling depth is highly desirable and such a device is
currently being prototyped.


\section{Summary}
A Switched Capacitor Array (SCA) device has been developed in a
0.25$\mu$m CMOS process with a 3dB analog bandwidth of almost a
Giga-Hertz, capable of being sampled at many GSa/s, or well above
Nyquist minimum. Sampling is performed at low power and the entire array of 9
channels by 260 samples can be digitized to 12-bits of resolution and
read out within 50$\mu $s.  With calibration excellent time and sample
voltage resolution have been obtained over a large range of
temperature and sampling speeds.


\section{Acknowledgements}
This work was supported by the National Aeronautics and Space
Administration (ROSS Program), the Department of Energy (HEP Division)
University of Hawaii base program support as well as support from the
Advanced Detector Research program.


\end{document}